\documentclass{article}
\usepackage{graphics}

\def\kbras{{_k \langle }}
\def\dint{\int \hspace{-0,25cm} \int}

\begin{document}

\title{\bf{On the construction of generalized Grassmann representatives of state
vectors}}

\author{{\bf{M. El Baz}}\thanks{moreagl@yahoo.co.uk} \hspace{0.2cm} and \hspace{0.2cm} {\bf{Y. Hassouni}}\thanks{y-hassou@fsr.ac.ma} }

\date{}
\vspace{-0.6cm} \maketitle{ \begin{center} \it Facult\'e des
sciences, D\'epartement de Physique, LPT, \linebreak Av. Ibn
Battouta, B.P. 1014, Agdal, Rabat, Morocco \end{center}}

\vspace{0.5cm}

\begin{flushright}
{published in { \it J. Phys. A37 (2004) 4361}}
\end{flushright}

\vspace{0.5cm}

\begin{abstract}
Generalized $Z_k$-graded Grassmann variables are used to label
coherent states related to the nilpotent representation of the
q-oscillator of Biedenharn and Macfarlane when the deformation
parameter is a root of unity. These states are then used to
construct generalized Grassmann representatives of state vectors.
\end{abstract}

\vspace{0.5cm}

Recently in \cite{mine3} we have constructed coherent states for
$k$-fermions using Kerner's \cite{kerner1} $Z_3$-graded extension
of the Grassmann variables \cite{berezin}. These results were
obtained in the case where the deformation parameter  is a
primitive cubic root of unity, i.e., $k=3$. In order to obtain
similar results in the generic case (by generic we mean here that
$q$, the deformation parameter, is an arbitrary $k^{th}$ root of
unity , i.e., for an arbitrary positive integer $k$) one should
use $Z_k$-graded generalizations of the Grassmann variables.
Unfortunately, up to our knowledge, such structures have not been
yet constructed in the spirit of Kerner's variables. There exist
however in the literature another point of view and other
generalized $Z_k$-graded Grassmann variables \cite{majid}.

In this letter we investigate on the use of these latter variables
for the description of $k$-fermions. Namely, we will construct
$k$-fermionic coherent states labeled by $Z_k$-graded Grassmann
variables. The coherent states will be used to derive a space of
($Z_k$-graded) Grassmann representatives in which state vectors
are represented as "holomorphic" functions of the Grassmann
variable. There exist many deformations of the harmonic oscillator
which, for some values of the deformation parameter, give rise to
$k$-fermions. In this letter we will illustrate the construction
using the oscillator deformation of Biedenharn \cite{biedenharn}
and Macfarlane \cite{macfarlane}. This $q$-oscillator is described
by the operators $\{N, a, a^+\}$ with the following relations
\begin{equation}
aa^+-q^{\pm 1} a^+a = q^{\mp N} \; , \;\;\;\;\; [N,a]=-a \; ,
\;\;\; [N,a^+]= a^+ \; . \label{BM}
\end{equation}

When $q$ is a $k^{th}$-root of unity, i.e., $q=\exp{({2\pi i \over
k})}$, this oscillator admits nilpotent representations (see e.g.
\cite{biedenharnbook}) in which we have
\begin{equation}
(a)^k = (a^+)^k = 0 \; . \label{BMroot}
\end{equation}

Note the following useful formulas, valid when $q$ is
$k^{th}$-root of unity,
\begin{equation}
q^k =1 \; , \; \; \; \bar q = q^{-1} = q^{k-1} \; , \;\;\; 1+ q+
\ldots q^{k-1} = 0 \; .
\end{equation}

The Fock space of this representation is finite dimensional of
dimension $k$, with basis
\begin{equation}
 \{ |n\rangle =
{(a^+)^n \over \big([n]_q\big)^{1\over 2}}  \, , \;\; a|0\rangle
=0 \, , \;\;  n= 0,1, ..., k \} \label{kbasis}
\end{equation}
 on which the different operators act as follows
\begin{eqnarray}
a|n\rangle &=& \big([n]_q\big)^{1\over 2} \; |n-1\rangle \nonumber
\\ a^+ |n\rangle &=& \big([n+1]_q \big)^{1\over 2} \;  |n+1\rangle
\label{qFock}
\\ N|n\rangle &=& n \;|n\rangle \nonumber
\end{eqnarray}

\vspace{0,5cm}

Next we remind the generalized $Z_k$-graded Grassmann variables
\cite{majid}. They obey the following $q$-commutation relations:
\begin{equation}
\begin{array}{cclc}
\xi_i \xi_j &=& q \xi_j \xi_i \; \;\;\;\; & \cr &&& i,j=
1,2,\ldots , d. \;  \; \; i<j \cr (\xi_i)^k &=& 0
\end{array} \label{theta}
\end{equation}
$d$ being the dimension of the Grassmann algebra, and $q$ is the
same as the deformation parameter of the $q$-oscillators we are
using.

The $\xi$'s are attributed the grade 1 and one introduces $p-1$
grade variables which are hermitian conjugates of the $\xi$'s:
$(\xi)^\dag = \bar\xi$, and obey similar relations:
\begin{equation}
\begin{array}{cclc}
\bar\xi_i \bar\xi_j &=& q \bar\xi_j \bar\xi_i \; \;\;\;\; & \cr
&&&  i<j \cr (\bar\xi_i)^k &=& 0\; .
\end{array}\label{bar}
\end{equation}

We have also the following relations between the two sectors
\begin{equation}
\begin{array}{cclc}
\xi_i \bar\xi_j &=& \bar q \bar\xi_j \xi_i \; \;\;\;\; & \cr &&&
i<j \cr \bar\xi_i \xi_j &=& \bar q \xi_j \bar\xi_i\, . \; \;\;\;\;
& \cr
\end{array}\label{bartheta}
\end{equation}

All these relations (\ref{theta}, \ref{bar}, \ref{bartheta}) can
be written in a condensed form:
\begin{equation}
\alpha_i \beta_j = q^{ab} \beta_j \alpha_i \;\;\;\; i<j
\end{equation}
where $a$ and $b$ are respectively the grades of $\alpha$ and
$\beta$; ($\alpha ,\,  \beta = \xi , \, \bar\xi$).

In the following we will confine our selves to the one dimensional
case, i.e., $d=1$; which means that we will drop the indices and
deal with one $\xi$, one $\bar\xi$ and the one-mode q-oscillator
(\ref{BM}, \ref{BMroot}).

We will also need the following rules of integration \cite{majid},
which are generalizations of the Berezin rules for the ordinary
Grassmann variable, \cite{berezin}:
\begin{equation}
\int d\alpha \; \alpha^n = \delta_{n,k-1} \; . \label{kBerezinint}
\end{equation}
where $\alpha = \xi  , \; \bar\xi$ and $ n$ is any positive
integer. And we have the following relations
\begin{equation}
\begin{array}{cclcrcl}
\xi d\bar\xi &=& q \; d\bar\xi \xi & \;\;\;\;\;\;\;\;\;\;\;\; &
\bar\xi d\xi &=& q \; d\xi \bar\xi \cr \xi d\xi &=& \bar q \; d\xi
\xi & \;\;\;\;\;\;\;\;\;\;\;\; & \bar\xi d\bar\xi &=& \bar q \;
d\bar\xi \bar\xi \cr d\xi d\bar\xi &=& \bar q \; d\bar\xi d\xi\, .
&& \cr
\end{array} \label{dtheta}
\end{equation}
Note that these rules (\ref{kBerezinint}, \ref{dtheta}) allow to
compute the integral of any function over the Grassmann algebra
since any such function is written as a finite power series in
$\xi$ and $\bar\xi$:
\begin{equation}
f(\xi, \bar\xi) = \sum_{i,j =0}^{k-1}C_{i,j} \xi^i {\bar\xi}^j \,
.
\end{equation}

In the fermionic case $(k=2)$ we know that $\xi$ and $\bar\xi$
anticommute with the fermionic creation and annihilation
operators. Thus, in the generic case (arbitrary $k$), rather than
assuming that the Grassmann variables commute with the
$k$-fermionic operators as is usually done in the literature (see
e.g. \cite{daoud}), one should assume $q$-commutation relations
between $\xi$, $\bar\xi$ and $a$, $a^+$:
\begin{equation}
\begin{array}{rclcrcl}
\xi a^+ &=& q \; a^+ \xi & \;\;\;\;\;\;\;\;\;\;\;\; &\xi a &=&
\bar q \; a \xi \cr  \bar\xi a^+ &=& \bar q \; a^+ \bar\xi &
\;\;\;\;\;\;\;\;\;\;\;\; &
 \bar\xi a &=& q \; a \bar\xi \; . \cr
\end{array} \label{atheta}
\end{equation}

\vspace{0,5cm}

In analogy with the fermionic case and the $Z_3$-graded case
considered in \cite{mine3}, where coherent states are constructed
by acting on the vacuum $|0\rangle$ with the exponential function
(in the former case) and $q$-exponential (in the latter), coherent
states for $k$-fermions are obtained similarly:
\begin{equation}
|\xi\rangle_k = \sum_{n=0}^{k-1} {(a^+ \xi)^n \over [n]_q!} \;
|0\rangle \; := \; \hbox{exp}_q(a^+ \xi) \; |0\rangle \; .
\label{kCS}
\end{equation}

Using (\ref{atheta}) one can find the following relation
\begin{equation}
(a^+ \xi )^n  = {\bar q}^{{n(n+1) \over 2}} \, \xi^n (a^+)^n
\end{equation}
which allows us to expand the coherent states (\ref{kCS}) in the
basis (\ref{kbasis}) with Grassmann coefficients:
\begin{equation}
|\xi \rangle_k = \sum_{n=0}^{k-1} {\bar q}^{n(n+1) \over 2} \;
{\xi^n \over \big([n]_q!\big)^{1\over 2}} \; |n\rangle \; .
\label{kCS2}
\end{equation}

Next we will prove that, as any harmonic oscillator coherent
state, these states are eigenstates of the annihilation operator
with $\xi$ as eigenvalue:

 {\samepage \begin{eqnarray} a |\xi
\rangle_k &=& \sum_{n=0}^{k-1} {\bar q}^{n(n+1) \over 2} \; {a
\xi^n \over \big([n]_q!\big)^{1\over 2}} \; |n\rangle \nonumber
\\ &=& \sum_{n=0}^{k-1} {\bar q}^{n(n-1) \over 2} \; { \xi^n a
\over \big([n]_q!\big)^{1\over 2}} \; |n\rangle \nonumber \\ &=&
\sum_{n=1}^{k-1} {\bar q}^{n(n-1) \over 2} \; { \xi^n \over
\big([n-1]_q!\big)^{1\over 2}} \; |n-1\rangle \; ,
\end{eqnarray}}
where we have used (\ref{atheta}) to obtain the second equality
and $\{ a|n\rangle = [n]_q^{1\over 2} |n-1\rangle , \; a|0\rangle
=0 \}$ for the third one. Then making the change $\{m= n-1 \}$ and
the nilpotency of $\xi$: $\xi^k =0$, we obtain the desired result:
\begin{eqnarray}
a |\xi \rangle_k &=& \sum_{m=0}^{k-1} {\bar q}^{m(m+1)\over 2} \;
{\xi^{m+1} \over \big([m]_q!\big)^{1\over 2}} \; |m\rangle
\nonumber
\\ &=& \xi |\xi \rangle_k \; .
\end{eqnarray}

Coherent states are usually not mutually orthogonal, this is also
the case for the coherent states (\ref{kCS2}); let us compute the
scalar product of two such states:
\begin{eqnarray}
{\kbras}\xi_2 |\xi_1\rangle_k &=& \sum_{n,m=0}^{k-1} {\bar
q}^{n(n+1)\over 2} q^{m(m+1) \over 2} \; {1 \over
\big([n]_q!\big)^{1\over 2} \big([m]_q!\big)^{1\over 2}} \langle
m|{\bar\xi}_2^m \xi_1^n |n\rangle \nonumber \\ &=&
\sum_{n=0}^{k-1} \; {{\bar\xi _2^n} \xi_1^n \over [n]_q!}
\label{koverlap}
\end{eqnarray}
where we have used the orthonormality of the basis (\ref{kbasis}):
$\langle m|n\rangle = \delta_{m,n}$. We make use of the equality
\begin{equation}
(\bar\xi \xi )^n = {\bar q}^{n(n-1)\over 2} \; {\bar\xi}^n \xi^n
\end{equation}
to write the scalar product in the form
\begin{equation}
{\kbras}\xi_2|\xi_1\rangle_k = \sum_{n=0}^{k-1} q^{n(n-1) \over 2}
\; {(\bar\xi_2 \xi_1 )^n \over [n]_q!} \; . \label{koverlap2}
\end{equation}
Then using the following relations between the two box functions
$[n]_q$ and $\{n\}_Q = {1-Q^n \over 1-Q}$ see e.g. \cite{arik}
(note that $\{n\}_{\bar q} \neq \{n\}_q$ whereas $[n]_{\bar q} =
[n]_q$):
\begin{equation}
[n]_q = [n]_{\bar q} = {\bar q}^{n-1} \; \{n\}_{q^2} = q^{n-1} \;
\{n\}_{{\bar q}^2} \; ,
\end{equation}
the corresponding $q$-factorials are then related by the
following:
\begin{equation}
[n]_q! = q^{n(n-1) \over 2} \; \{n\}_{{\bar q}^2}! \; .
\end{equation}
This permits us to rewrite the scalar product (\ref{koverlap2}) as
follows:
\begin{eqnarray}
{\kbras}\xi_2 |\xi_1\rangle_k &=& \sum_{n=0}^{k-1} { (\bar\xi_2
\xi_1)^n \over \{n\}_{{\bar q}^2}!} \nonumber \\ &:=& E_{{\bar
q}^2}(\bar\xi_2 \xi_1)\; ,
\end{eqnarray}
where we have used the following $q$-exponential \cite{arik}
\begin{equation}
E_q(\alpha) = \sum_{n=0}^{k}{\alpha^n \over \{n\}_q!} \; ; \;\;
\{n\}_q! = \{1\}_q \ldots \{n\}_q \; ;\;\; \{0\}_q!:=1 \;.
\end{equation}

\vspace{0,5cm}

One of the most important defining properties of coherent states
is that they provide a resolution of unity \cite{klauder}. Thus in
order to prove that the states (\ref{kCS2}) are indeed coherent
states we shall show that they allow a resolution of unity. We
shall look for such a resolution in the form
\begin{equation}
\dint d\bar\xi d\xi \; \omega (\bar\xi \xi) \;
|\xi\rangle_k{\kbras}\xi| = I \; , \label{kRI}
\end{equation}
where the weight function is written as follows
\begin{equation}
\omega(\bar\xi \xi) = \sum_{n=0}^{k-1} \; c_n \; \xi^n {\bar\xi}^n
\; . \label{weight1}
\end{equation}

We must compute the coefficients $c_n$ such that the equality in
(\ref{kRI}) holds.

Using (\ref{kCS2}) and (\ref{weight1}) the left hand side of
(\ref{kRI}) is written as follows
\begin{equation}
\dint \sum_{l,n,p=0}^{k-1} d\bar\xi d\xi \; c_n \xi^n {\bar\xi}^n
\; {\bar q}^{l(l+1)\over 2} q^{p(p+1)\over 2} \; {\xi^l
{\bar\xi}^p \over \big([l]_q!\big)^{1\over 2}
\big([p]_q!\big)^{1\over 2}} \; |l\rangle\langle p| \; .
\end{equation}
Then taking account of relations (\ref{bartheta}) and the
integration rules (\ref{kBerezinint}), it becomes
\begin{equation}
\dint d\bar\xi d\xi \; \sum_{l,n=0}^{k-1} {c_n q^{nl} \over
[l]_q!} \; \xi^{n+l} {\bar\xi}^{n+l} |l\rangle\langle l| \; .
\end{equation}

The Fock space basis being complete: $\displaystyle
\sum_{l=1}^{k-1} |l\rangle\langle l| = I$, now using
(\ref{kBerezinint}) we obtain the following constraints on the
coefficients $c_n$
\begin{equation}
I= \sum_{\stackrel{n,l=0}  {n+l = k-1}}^{k-1} {c_n q^{nl} \over
[l]_q!} \dint d\bar\xi d\xi \xi^{n+l} {\bar\xi}^{n+l} \;
|l\rangle\langle l| \, ,
\end{equation}
i.e.,
\begin{equation}
{c_n q^{nl} \over [l]_q!} = 1 \;\; \hbox{and} \;\;\; n+l = k-1 \;.
\end{equation}

We have thus found the coefficients $c_n$, for which the equality
(\ref{kRI}) holds, to be
\begin{equation}
\begin{array}{ccl}
c_n &=& {\bar q}^{n(k-n-1)} [k-n-1]! \cr &=& q^{n(n+1)} [k-n-1]!
\cr
\end{array}
\end{equation}

The weight function appearing in the resolution of unity
(\ref{kRI}) is therefore given by
\begin{eqnarray}
\omega(\bar\xi \xi) &=& \sum_{n=0}^{k-1} c_n \xi^n {\bar\xi}^n
\nonumber \\ &=& \sum_{n=0}^{k-1} q^{n(n+1)} [k-n-1]_q! \xi^n
{\bar\xi}^n \nonumber \\ &=& \sum_{n=0}^{k-1} q^{n(n+1) \over 2}
[k-n-1]_q! (\bar\xi \xi)^n\; .
\end{eqnarray}

\vspace{0,5cm}

Coherent states allow in general the mapping of vectors of the
underlying Hilbert space into holomorphic functions (Bargmann-Fock
space), when they (the coherent states) are parameterized by a
complex variable. In the case where they are parameterized by
Grassmann variables  (fermionic case \cite{klauder, ohnuki} or
$k$-fermionic case \cite{mine3}), coherent states allow the
mapping into a space of Grassmann representatives. In all cases it
is the resolution of unity that permits this mapping.

However, when dealing with generalized Grassmann variables, it was
shown in \cite{mine3} that it is more convenient, and sometimes
essential, to use another form of the resolution of unity, namely
\begin{equation}
\dint |\xi\rangle_k \; d\bar\xi d\xi \; {\tilde\omega}(\bar\xi
\xi) \; {\kbras}\xi| = I \; , \label{kRI2}
\end{equation}
where the weight function $\tilde \omega$ is written in the same
form as $\omega$ in (\ref{weight1}), with {\it a priori} different
coefficients:
\begin{equation}
\tilde\omega (\bar\xi \xi) = \sum_{n=0}^{k-1} {\tilde c}_n \xi^n
{\bar\xi}^n \; .\label{baromega}
\end{equation}

We have deliberately given here the two forms (\ref{kRI},
\ref{kRI2})of the resolution of unity to insist on the fact that
in the case we are considering in this letter these two forms are
equivalent, whereas in \cite{mine3} (where we used the
$Z_3$-graded Grassmann variables of Kerner \cite{kerner1}) they
are not. By equivalent we mean that one can deduce one form from
the other. So, the coefficients $\tilde{c}_n$ in (\ref{baromega})
can be evaluated either by deducing them from the coefficients
$c_n$ directly or by explicitly performing the calculations as for
the $c_n$. In any case the result is the following:
\begin{equation}
\begin{array}{ccl}
{\tilde c}_n &=& [k-n-1]_q! \cr &=& {\bar q}^{n(n+1)} c_n \; .\cr
\end{array}
\end{equation}
These are the values for which the equality in (\ref{kRI2}) holds.

\vspace{0,5cm}

We are now in a position to construct the Grassmann
representatives of state vectors. Indeed, using the latter form of
the resolution of unity (\ref{kRI2}), any vector  $|\psi\rangle$
in the Hilbert space spanned by the basis $\{|n\rangle \, , \;
n=0,1,\ldots , k-1\}$ can be determined by its Grassmann
representative $\psi (\bar\xi) :=\,  {\kbras}\xi |\psi\rangle$:
\begin{equation}
|\psi\rangle = \dint |\xi\rangle_k d\bar\xi d\xi \;
{\tilde\omega}(\bar\xi \xi) \; \psi(\bar\xi) \; .
\end{equation}

In particular, the basis vectors are realized by the following
monomials in $\bar\xi$:
\begin{equation}
n(\bar\xi) := \, \langle \xi|n\rangle = {\bar q}^{n(n-1) \over 2}
{{\bar\xi}^n\over \big([n]_q!\big)^{1\over 2}} \; ,
\end{equation}
which, therefore, constitute a basis  of the space of {\it
polynomial} functions over the Grassmann algebra generated by
${\bar\xi}$ also called space of (generalized) Grassmann
representatives.

The scalar product in this realization space is obtained using
(\ref{kRI2}) and is given by
\begin{equation}
\langle \psi|\varphi\rangle = \dint {\bar\psi}(\xi) \; d\bar\xi
d\xi \; {\tilde\omega}(\bar\xi \xi) \; \varphi(\bar\xi) \; .
\end{equation}

We check that, in particular, the orthonormality of the basis
$\{|n\rangle \, ,\; n=0,1, \ldots , k-1\}$ is preserved under this
scalar product:
\begin{eqnarray}
\langle m|n\rangle &=& \dint {\bar m}(\xi) \; d\bar\xi d\xi \;
{\tilde\omega}(\bar\xi \xi) \; n(\bar\xi) \nonumber \\ &=& \dint
\sum_{l=0}^{k-1} {q^{m(m-1)\over 2} \over \big([m]_q!\big)^{1\over
2}} {{\bar q}^{n(n-1) \over 2} \over \big([n]_q!\big)^{1\over 2}}
\; {\tilde c}_l \xi^m \; d\bar\xi d\xi \; \xi^l \bar\xi^l
\bar\xi^n \nonumber \\ &=& \dint \sum_{i=0}^{k-1} {{\tilde c}_l
\over [n]_q!} \delta_{m,n} \; d\bar\xi d\xi \; \xi^{l+n}
{\bar\xi}^{l+n} \nonumber \\ &=& \sum_{i=0}^{k-1} {{\tilde c}_l
\over [n]_q!} \delta_{m,n} \; \delta_{l,k-n-1} \nonumber \\ &=&
{{\tilde c}_{k-n-1} \over [n]_q!}\delta_{m,n} \nonumber \\ &=&
\delta_{m,n} \; .
\end{eqnarray}
In the last equality, we have used the fact that
\begin{equation}
{\tilde c}_{k-n-1} = [n]_q! \; .
\end{equation}

\vspace{0,5cm}

Moreover, if we denote by $\partial_{\bar\xi}$ the (ordinary)
derivative operator:
\begin{equation}
\partial_{\bar\xi} \; {\bar\xi}^m = m {\bar\xi}^{m-1}
\end{equation}
and $D_{\bar\xi}$ the deformed derivative operator:
\begin{equation}
D_{\bar\xi} \; {\bar\xi}^m = [m]_q \; {\bar\xi}^{m-1} \; ;
\end{equation}
actually this last operator is defined as follows
\begin{equation}
D_{\bar\xi} \; f(\bar\xi) = {f(q\bar\xi) - f(q^{-1} \bar\xi) \over
(q-q^{-1})\bar\xi} \; ;
\end{equation}
then one can easily realize the annihilation and creation operator
as differential operators, acting on the space of Grassmann
representatives, as follows
\begin{eqnarray}
a &\longrightarrow & q^{\bar\xi \partial_{\bar\xi}} \; D_{\bar\xi}
\\ a^+ &\longrightarrow & {\bar\xi} \; {\bar q}^{\bar\xi
\partial_{\bar\xi}}
\end{eqnarray}

As a matter of facts the differential operator $\bar\xi
\partial_{\bar\xi}$ appearing in these expressions is the
realization of the number operator $N$.

\vspace{0,5cm}

It is obvious that in order to obtain a realization in terms of
$\xi$, rather than $\bar\xi$, one should use coherent states
labeled by $\bar\xi$ (rather than $\xi$), i.e., $|\bar\xi\rangle$.

\vspace{0,5cm}

In summary, we have constructed coherent states, which are labeled
by $Z_k$-graded Grassmann variables, relative to the nilpotent
representation of the $q$-oscillator of Biedenharn and Macfarlane
when $q$ (the parameter of deformation) is a $k^{th}$ root of
unity. We have then showed that these coherent states map the
Hilbert space of the representation of the $q$-oscillator into a
space of some {\it polynomial} functions over the Grassmann
variables, i.e., the space of Grassmann representatives. The
different operators of the $q$-oscillator algebra act on this
space as differential operators involving both ordinary and
deformed derivatives.

At the same footstep, the coherent states constructed in this
space can be used to obtain Grassmann representatives of any
operator acting on the Hilbert space, using some sort of {\it
Grassmann covariant symbol} of the operator. This should prove
useful for the construction of a path integral formalism with
$Z_k$-graded Grassmann variables.

It is worth to mention that our starting point here for the
construction of $k$-fermionic coherent states was the assumption
of the non-commutativity of the Grassmann variables with the
$k$-fermionic operators (\ref{atheta}), instead of assuming
commutativity as is generally the case in the literature
\cite{daoud}. We would like to argue here that this last point of
view (i.e., commutativity instead of (\ref{atheta})) does not seem
to be consistent. In fact, one of the features of $k$-fermionic
oscillators is that they reduce to the usual fermionic one, when
the deformation parameter takes a special value. One expects this
property to be preserved when constructing further $k$-fermionic
structures, such as coherent states. Moreover, taking into account
that in the usual fermionic case the Grassmann variables do not
commute with the fermionic operators but rather anticommute, it is
easy to check that this behaviour is recovered in the fermionic
limit by adopting the point of view presented in this letter
(i.e., $k=2$) from (\ref{atheta})) but not by assuming
commutativity. Therefore, it is easy to convince oneself that the
point of view generally adopted in the literature, even though
rendering the calculations simpler, is not consistent since it
fails to encompass the fermionic behaviour, which we are supposed
to generalize by considering $k$-fermionic structures.

Finally, let us note that we have used the $q$-oscillator of
Biedenharn and Macfarlane in this letter, the construction is
however easily extended to other $q$-oscillators giving rise to
$k$-fermions.

\section*{Acknowledgments}

Y. H. acknowledges financial support from CBPF.

\end{document}